\documentclass[14pt]{article}

\usepackage{arxiv}
\usepackage[utf8]{inputenc} 
\usepackage[T1]{fontenc}    
\usepackage{hyperref}       
\usepackage{url}            
\usepackage{booktabs}       
\usepackage{amsfonts}       
\usepackage{nicefrac}       
\usepackage{microtype}      
\usepackage{lipsum}		
\usepackage[super,compress]{cite}
\usepackage{graphicx}
\usepackage{dcolumn}
\usepackage{bm}
\usepackage{array}
\usepackage{tabularx}
\usepackage[english]{babel}
\usepackage[nottoc]{tocbibind}

\title{The Effect of Dark Matter on Stars at the Galactic Center: The Paradox of Youth Problem}

\author{
  Ebrahim Hassani \\
  Department of Physics, Faculty of Sciences, \\
  University of Birjand, Birjand, Iran\\
  \texttt{ebrahim.hassani@birjand.ac.ir} \\
   \And
  Reza Pazhouhesh \\
	Department of Physics, Faculty of Sciences, \\
	University of Birjand, Birjand, Iran\\
  \texttt{rpazhouhesh@birjand.ac.ir} \\
   \And
  Hossein Ebadi \\
Department of Theoretical Physics and Astrophysics \\
University of Tabriz, PO Box 51664, Tabriz, Iran\\
   \texttt{hosseinebadi@tabrizu.ac.ir} \\
}

\begin{document}
	
\maketitle

\tableofcontents

\begin{abstract}
	\begin{description}
		\item[Abstract]
		Stars that evolve near the Galactic massive black hole show strange behaviors. The spectroscopic features of these stars show that they must be old. But their luminosities are much higher than the amounts that are predicted by the current stellar evolutionary models, which means that they must be active and young stars. In fact this group of stars shows signatures of old and young stars, simultaneously. This is a paradox known as the "paradox of youth problem" (PYP). Some people tried to solve the PYP without supposing dark matter effects on stars. But, in this work, we implemented Weakly Interacting Massive Particles (WIMPs) annihilation as a new source of energy inside such stars. This implementation is logical for stars that evolve at high dark matter density environments. The new source of energy causes stars to follow different evolutionary paths on the H-R diagram in comparison with classical stellar evolutionary models. Increasing dark matter density in stellar evolutionary simulations causes the deviations from the standard H-R diagrams becomes more pronounced. By investigating the effects of WIMPs density on stellar structures and evolutions, we concluded that by considering dark matter effects on stars at the Galactic center, it is possible to solve the PYP. In addition to dark matter effect, complete solutions to PYP must consider all extreme and unique physical conditions that are present near the Galactic massive black hole.
	\end{description}
\end{abstract}

\section{\label{sec:level1}Introduction}

Rotation curves of galaxies show that dark matter (DM) is distributed throughout almost all galaxies \cite{cosmology_weinberg}. But DM density is not uniform inside galaxies. For instance, near the solar system it is approximated to be about $\sim 8.16 \times 10 ^{-3} \: M_{\odot}  pc^{-3}$ \cite{BERTONE_2005}. While, this is about $\sim 10 ^{15} M_{\odot} pc ^{-3}$ near the Galactic massive black hole (MBH) \cite{BERTONE_2005_b} . So, we can infer that almost all stars that evolve inside galaxies immersed in the DM hallo. Then, it is logical to suppose that DM must affect stars structure and evolutionary tracks inside galaxies.

In addition to higher DM density near the Galactic MBH, there are also other extreme and unique physical conditions that affect the structure and evolutionary track of stars. For example: a) density of stars are high near the Galactic MBH in comparison with other places in the Galaxy. For instance, in the distance about 0.1 parsec from the Galactic MBH, stars density is about $10^{8} \: M_{\odot} \: pc^{-3}$, which is $10^{9}$ times bigger than the amount near the solar system \cite{ALEXANDER200565}. b) Tidal effects are very high near the Galactic MBH \cite{ALEXANDER200565}.

Observational evidences show that the evolutionary track of stars near the Galactic MBH must be different from other stars in the Galaxy \cite{Demarque2007, Ghez2003}. Spectroscopic features of these stars show that they are evolved stars and must be old. Besides, they are much luminous than the amounts that are predicted by the current stellar evolutionary models, then they must be young and active. This is a paradox known as the "Paradox of Youth Problem" (PYP) \cite{Ghez2003}. PYP arises from the fact that current stellar evolutionary models (CSEM) do not consider the effects of extreme physical conditions that are present near the Galactic MBH. Although, there are attempts to solve the PYP via CSEMs \cite{Demarque2007}, but, as will be discussed more in next sections, CSEMs are unable to interpret the strange behavior of stars near the Galactic MBH.

PYP can be solved by considering DM annihilation as a new source of energy inside stars. In fact, we showed that by considering DM effects inside stars, stars will show sings of young and old stars simultaneously. Then the PYP is solved.

For the first time, Steigman et al. had studied the DM effect on the sun as a possible solution to the solar neutrino problem \cite{Steigman_1978} . Since then many studies had conducted to study possible impacts of the DM on the sun \cite{Gould_1987, Spergel_1985, Gilliland_1986, Press_1985, Turck_Chi_2012, Cumberbatch_2010, Casanellas_2014, Adrian_Martinez_2013, Lopes_2002} . Later studies (see e.g. \cite{Salati_1989} and \cite{Bouquet_1989})  explored the inherent impacts that DM can produce in the stellar structures and evolutions. Subsequent series of studies tried to predict and find the possible observational evidences that DM can have on the stellar structures and evolutions \cite{Dearborn_1990, Taoso_2008, Scott_2008, Fairbairn_2008, Zentner_2011, Lopes2011, Casanellas_2009, Martins_2017} . In addition to normal stars, the effect of dark matter on compact stars (neutron stars and white dwarfs) was also investigated. For instance, it is possible to constrain DM properties using compact stars \cite{Bertone_2008, Kouvaris_2010} or investigate DM properties using pulsars at the Galactic center \cite{Bramante_2014} . In addition, annihilation of WIMPs inside old neutron stars can flatter out their temperature \cite{Kouvaris_2008} .

In addition to the PYP, several other clues show that DM may affect stellar structures and evolutionary tracks inside galaxies. For example: a) Cold DM at the center of dwarf galaxies can be heated-up by stars and then move around. Hence, at a fixed dark matter halo mass, dwarfs with a higher stellar mass will have a lower central dark matter density \cite{Read2019}. b) Several solutions are proposed to solve solar neutrino discrepancy. One of them emphasizes on the deficiency of the solar standard model. Solar standard model does not consider the effects of DM in its assumptions \cite{Stix2004}. But, this problem can be solved by considering DM effects in solar standard model \cite{Adrian_Martinez_2013}.

In this work, after investigating the effects of DM density on stars evolutions and structures, we concluded that PYP can be solved by considering DM annihilation as a new source of energy inside stars. In fact, we showed that by considering DM effects inside stars, stars can show signs of young and old stars simultaneously. Then, the PYP is solved.

From many particle physics candidates, we selected Weakly Interacting Massive Particles (WIMPs) which have the most agreement with the current known physical properties of DM \cite{BERTONE_2005, Bertone2018}. In this work we selected $ M_{x} = 50 \: Gev $ for WIMPs mass, which lies within the accepted mass range for WIMPs \cite{BERTONE_2005}.

In section \ref{sec:level3}, effects of DM density on the evolutionary track of some stars was investigated. In section \ref{sec:sec2-1}, effect of DM density on the capture rate by stars was investigated. In section \ref{sec:sec2-2} effect of DM density on the total amount of WIMPs that can be captured by a star was studied. In section \ref{sec:sec2-3} effect of DM density on the total luminosity that can be produced by WIMP self-annihilation was calculated. In section \ref{sec:sec2-4} effect of DM density on the distribution of WIMPs inside stars was studied. Section \ref{sec:level4} devoted to discussions and conclusions.

\section{\label{sec:level3}Effect of Dark Matter Density on Stellar Evolutions}

\begin{figure*}
	\includegraphics[width=\textwidth]{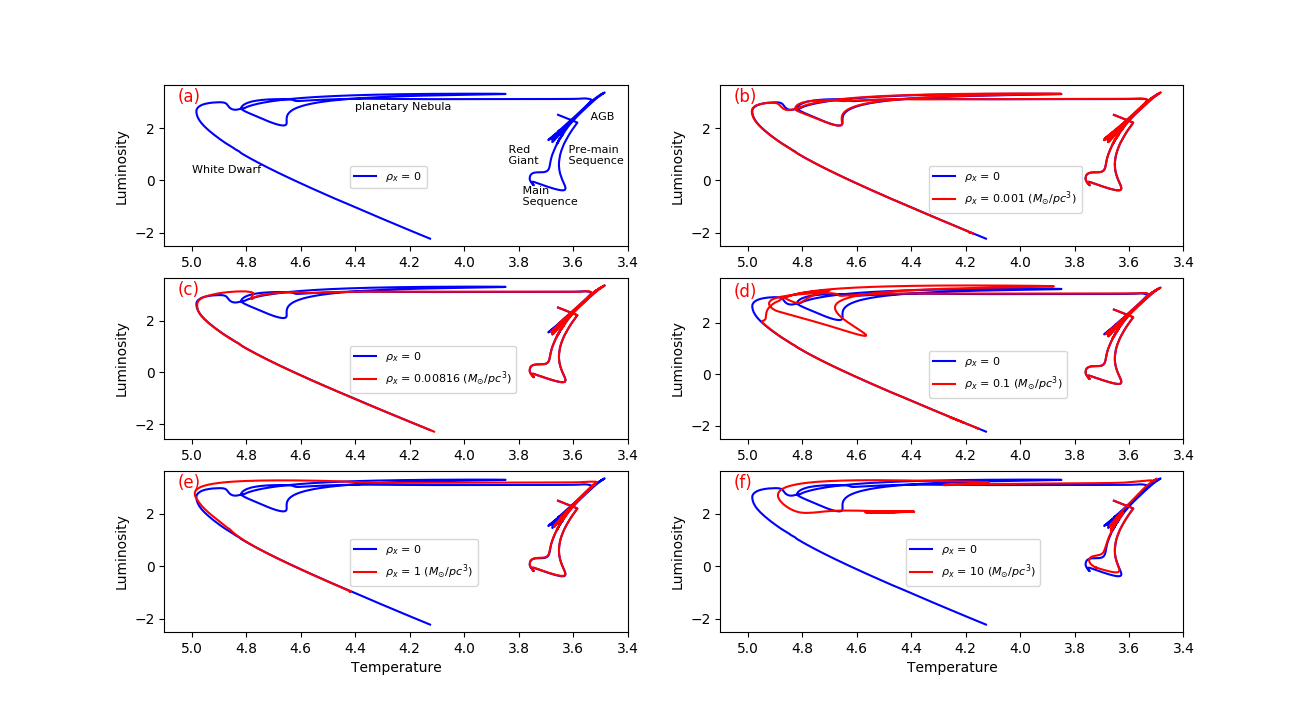}
	\caption{\label{fig:solar_evol} (color online). Evolution of solar mass star on H-R diagram in different DM density environments. In all diagrams blue lines are for $ \rho_{x} = 0 $ state and red lines are for $ \rho_{x} \ne 0 $ state. By increasing DM density, evolutionary path of the star deviates from standard stellar evolutionary model. Part a) is for $ \rho_{x} = 0 $ state, b) for $ \rho_{x} = 0.001  M_{\odot} / pc^{3} $ state, c) for $ \rho_{x} = 8.16 \times 10 ^{-3} M_{\odot} / pc^{3} $ state, d) for $ \rho_{x} = 0.1 M_{\odot} / pc^{3} $ state, e) for $ \rho_{x} = 1  M_{\odot} / pc^{3} $ state, f) for $ \rho_{x} = 10 M_{\odot} / pc^{3} $ state}
\end{figure*}

In this section, as an example, effect of DM density on the evolutionary track of some stars is investigated. We implemented equation \ref{eq:L_x} (see section \ref{sec:sec2-4}) as a new source of energy in MESA code and then evolved the star (run the MESA code). MESA is a free and open source code which can simulate stars from very low mass to very high mass ($ \approx 10^{-3} - \: 10^{3} M_{\odot} $). The full capabilities of MESA are documented in its official instrument papers \cite{Paxton_2010, Paxton_2013, Paxton_2015, Paxton_2018}. In this work we used 10398 version of this code. The results are presented in figure \ref{fig:solar_evol} which demonstrates H-R diagram of a solar mass star that evolves in different DM density environments. In this figure blue lines represents a star without DM ($ \rho_{x} = 0 $) and red lines are for stars with $ \rho_{x} \ne 0 $ state. $ \rho_{x} $ is the density of DM at the position of the stars.

Figure \ref{fig:solar_evol}-a demonstrates the evolution of standard solar mass star ($ \rho_{x} = 0 $) in the H-R diagram. By paying attention to other parts of figure \ref{fig:solar_evol} (from figure \ref{fig:solar_evol}-a to figure \ref{fig:solar_evol}-f), we can see that by increasing DM density, the evolutionary path of the stars on the H-R diagram deviates more from the standard model.

As an example, figure \ref{fig:solar_evol}-c demonstrates the evolutionary path of a solar mass star in $ \rho_{x} = 8.16 \times 10 ^{-3} (M_{\odot} / pc^{3}) $ DM density environment. This amount is about the DM density near the sun position \cite{BERTONE_2005} . When star is in the main-sequence phase it does not deviate a lot from the solar standard model. But when it reaches the planetary-nebulae phase its evolutionary path deviates a lot from the standard model. The reason for this behavior is that when star is in the main-sequence phase, it is very active and most of its energy is acquired through hydrogen burning at the center. Then, the portion of the produced energy through WIMP-annihilation is insignificant in comparison to the baryonic matter energy production cycles. The story will change when the star reaches to the planetary nebula phase. In this phase, star exhausted almost all of its energy sources that comes from the baryonic matters. Then, in this phase, WIMPs annihilation are the predominant source of energy (or luminosity).

In other words, when stars exhaust the source of energy that comes from the baryonic matter, WIMP annihilation becomes the predominant energy source inside the stars. At this stage, although stars spectrum is similar to the evolved and old stars (due to presence of heavier elements in atmosphere of stars), but their luminosities are high (due to the constant luminosity that comes from the WIMPs annihilation) and gives active and young appearance to the star. In reality, stars near the Galactic MBH suffer from these features. They exhibit signatures of young and old stars, simultaneously. In simple words, the PYP may be solved by comparing known observational data with suitable theoretical models. In more denser DM environments, stars begin to deviate from standard model before they enter the planetary nebula phase. Then, they follow almost completely different evolutionary path on the H-R diagram. It means that, CSEMs are unable to interpret the evolution of stars in high DM density environments.

\begin{figure*}
	\includegraphics[width=\textwidth]{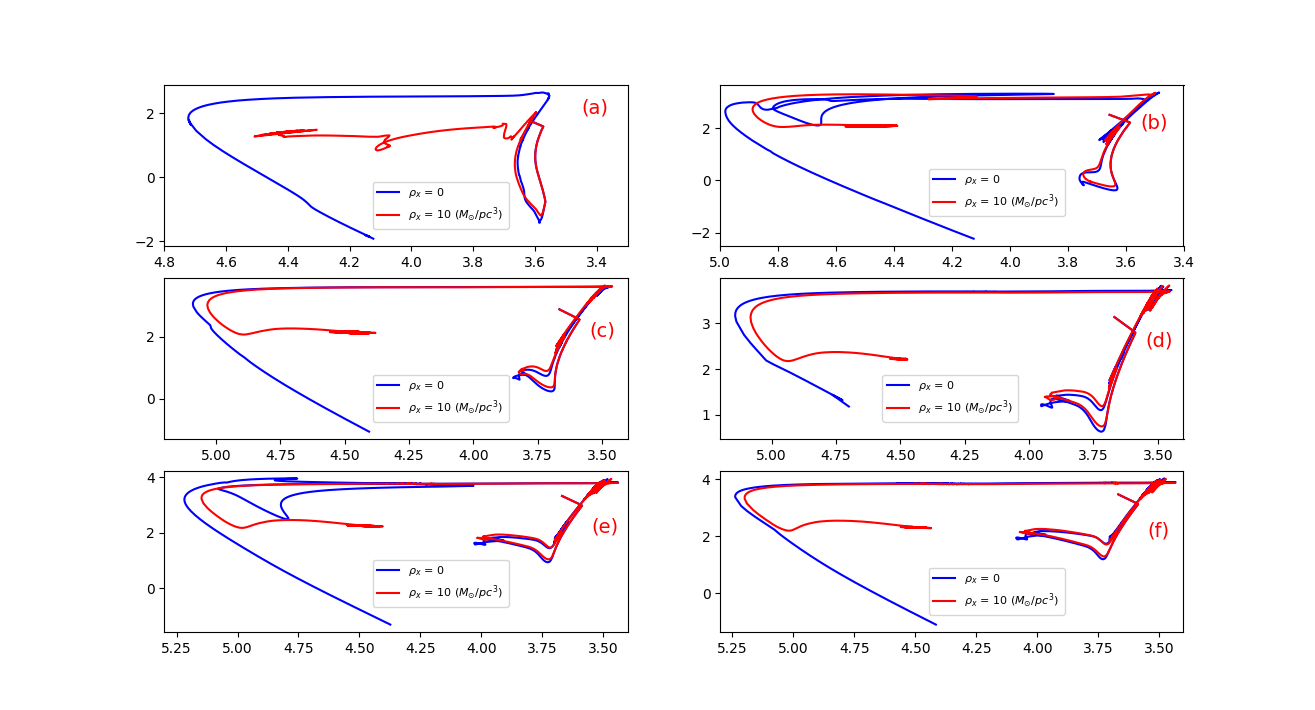}
	\caption{\label{fig:evolutionary_diagram} (color online). Theoretical H-R diagrams of stars in this study in two states: without DM (blue lines, $ \rho_{x} = 0 $) and with DM (red lines,  $ \rho_{x} = 10 M_{\odot} / pc^{3} $). Part (a) is for $0.5 M_{\odot}$ star, part (b) for $1.0 M_{\odot}$ star, part (c) for $1.5 M_{\odot}$ star, part (d) for $2.0 M_{\odot}$ star, part (e) for $2.5 M_{\odot}$ star and part (f) for $3.0 M_{\odot}$ star ($M_{\odot}$ is the mass of the sun). In all diagrams when stars reaches to the white dwarf (or neutron star for massive stars) phase, stars with DM (red lines) become more luminous than stars without DM (blue lines). At this phase, stars with DM look like active and young stars, while they are old and have spectroscopic features of old stars.}
\end{figure*}

Figure \ref{fig:evolutionary_diagram} illustrates similar results for stars with different stellar masses. During the first stages of the evolution of stars, they do not deviate a lot from the CSEM. This is because, at these phases, the amount of energy that comes from WIMPs annihilation is negligible in comparison with the amount of energy that comes from baryonic matter energy production cycles. At later phases (planetary nebulae and white dwarf or neutron star phases) the story will change and WIMP annihilation will become the dominant energy source inside stars. As an example, in figure \ref{fig:evolutionary_diagram}-b star with DM is almost $10^{4}$ times more luminous than that the model without DM, when it reaches to white dwarf phase. In addition, from figure \ref{fig:evolutionary_diagram} one can also infer that DM affects evolutionary tracks of low-mass stars more than high-mass ones. This result is in agrement with the results of the previous studies (e.g. see \cite{Casanellas_2014, Turck_Chi_2012, Casanellas_2009, Lopes2011, Zentner_2011, Salati_1989, Fabio_2012, Scott_2009, Bouquet_1989_b, Casanellas_2013 } ).

\section{\label{sec:level2}Effect of Dark Matter Density on Stellar Structures}

\subsection{\label{sec:sec2-1}Capture Rate}
According to definition, capture rate for a body (e.g. Earth, star, neutron star) is the number of DM particles that gravitationally bounding to that body in units of time. For the first time, capture rate of the sun calculated in paper \cite{Press_1985} . Then, Gould (1987) generalized capture rate relation for other round bodies (like Earth) \cite{Gould1987} . Bouquet and Salati obtained specific capture rate relation for main sequence stars \cite{Bouquet_1989} . In this work we used equation 42 of paper \cite{Bouquet_1989} for capture rate:

\begin{eqnarray} \label{eq:caprate}
C(M_{\star}, m_{x}, \sigma_{s}) \approx 1.2 \times 10^{30} s^{-1}\left(  m_{p} / m_{x} \right)
\left( M_{\star}/M_{\odot} \right)^{1.8}\nonumber\\
\times
\left(  \frac{300 kms^{-1}}{V_{halo}}  \right) \times \left(  \rho_{halo} / 0.01 M_{\odot} pc^{-3}  \right)\nonumber\\
\times Min \left\{   1 ,  \frac{\sigma_{s}}{\sigma _{c,M_{\odot}}}  \left[  \frac{M_{\odot_{}^{}}}{M_{\star}} \right]^{0.6}  \right\} \:\:\:\:\:\:\:\:\:\:\:\:\:\:
\end{eqnarray}

where $m_p$ is the mass of the protons, $m_x$ is the mass of the WIMPs, $M_{\star}$ is the mass of the star, $V_{halo}$ is the average speed of the WIMPs in the halo of the galaxy, $\rho _{halo}$ is the DM halo density at the location of the star, $\sigma _{c} \left( M _{\odot}\right) = 4 \times 10^{-36} cm^{2}$ is the average scattering cross section of the sun, $\sigma_{s}$ is the scattering cross section of the WIMPs from different elements in the star. In equation \ref{eq:caprate} all parameters are fixed except $\rho_{halo}$.

In equation \ref{eq:caprate} it is necessary to calculate $\sigma_{s}$ for different elements. We considered spin-dependent scattering cross section to be $\sigma_{\chi,SD} = 10 ^{-38} \: cm^{2}$ and spin-independent scattering cross section to be $\sigma_{\chi,SI} = 10 ^{-44} \: cm^{2}$. These are the maximum magnitudes that are determined through the experimental DM detection experiments \cite{Angle2008, Angle2008b} . As the portion of the heavy elements are very small in the main-sequence stars (in comparison with Hydrogen and Helium) we can neglect the scattering cross section for all elements heavier than Helium. In addition, scattering cross section of Helium atoms are very small in comparison to Hydrogen atoms. This is because of the fact that spin-dependent scattering cross section is $10^{6}$ times bigger than spin-independent one. These approximations are usually done in similar works \cite{Taoso_2008, Lopes2011} .

\begin{figure*}
	\includegraphics[width=\textwidth]{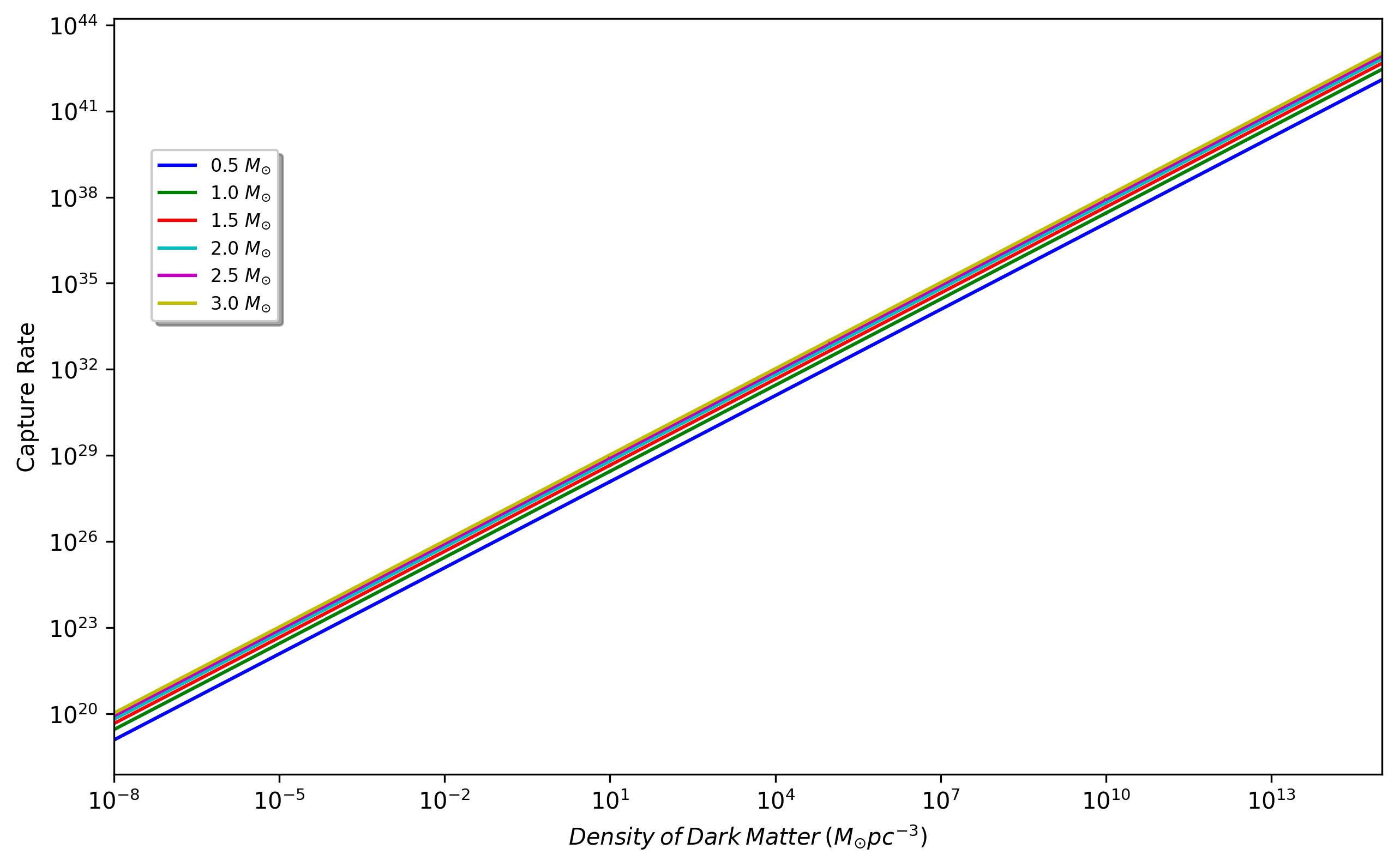}
	\caption{\label{fig:caprate} (color online). Capture rate for each star in table \ref{tab:stars_params} as function of DM density. Different colors represents different stellar masses. The massive the star is, then, the higher the capture rate is.}
\end{figure*}

Figure \ref{fig:caprate} shows the capture rate for stars with different stellar masses in different DM density environments. DM density varies from 0 to $ 10 ^{15} M_{\odot} pc ^{-3}$ for stars with masses $0.5 - 3.0 \: M_{\odot}$. As it is clear from Figure \ref{fig:caprate},  the capture rate for high-mass stars is higher than low-mass stars.

\subsection{\label{sec:sec2-2}Total Number of WIMPs}

\begin{figure*}
	\includegraphics[width=\textwidth]{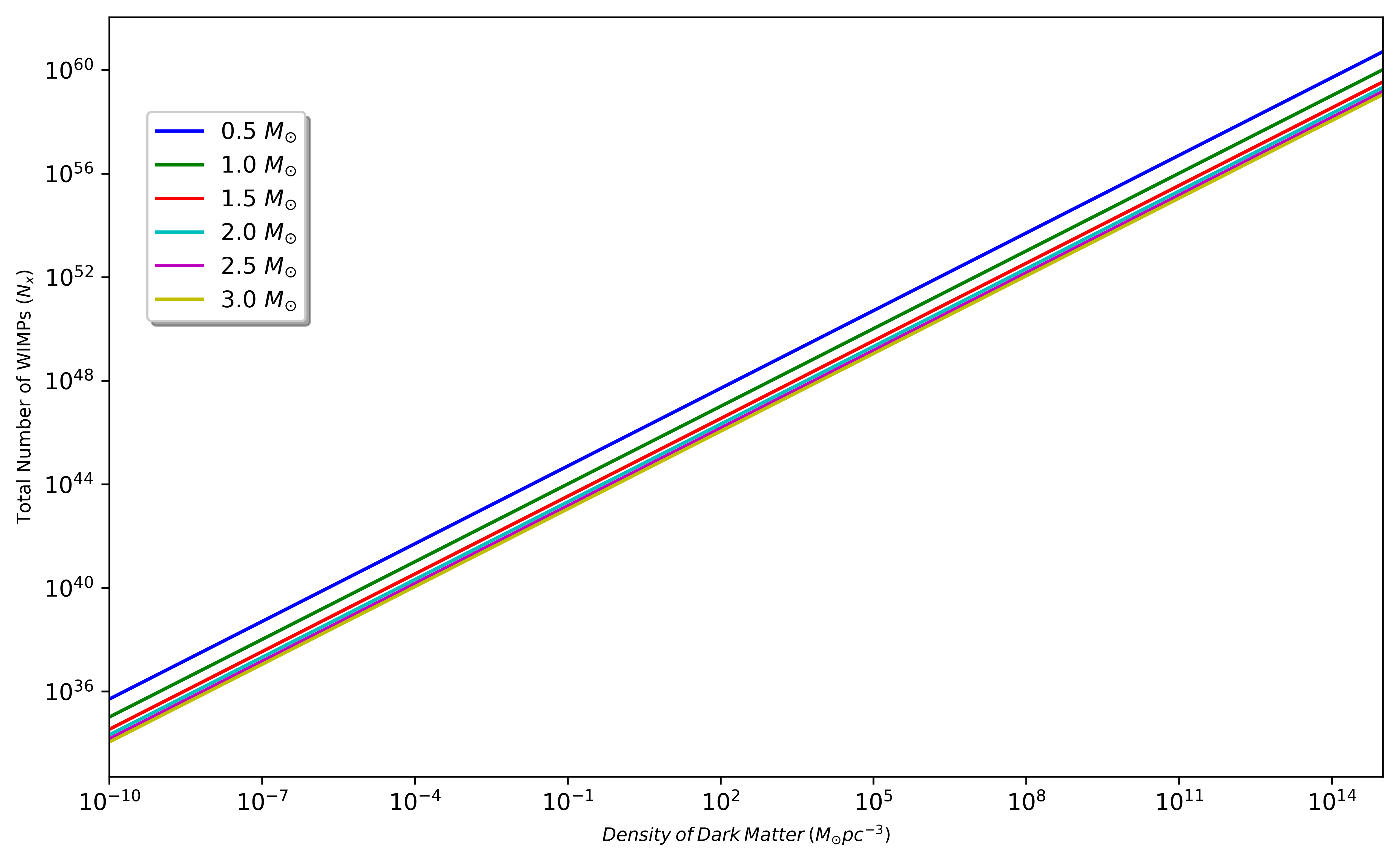}
	\caption{\label{fig:total_wimps} (color online). Total number of WIMPs that are captured by each star in table \ref{tab:stars_params} when they leave main-sequence. The lowest the mass of the star is, then, the highest the total number of WIMPs is. This is because, according to CSEMs, low-mass stars spend more time on the main-sequence phase in comparison to the high-mass ones (e.g. see age of stars in table \ref{tab:stars_params} and also see reference \cite{Prialnik2000}). Then, low-mass stars have enough time to capture more WIMPs.}
\end{figure*}

In order to calculate the total number of WIMPs that are captured by each star during the main-sequence phase, it is enough to multiply capture rate by $ t_{ms} $ ( $ t_{ms} $ is the main-sequence lifetime of a star). For extra explanation about $ t_{ms} $ and how to calculate it, see reference \cite{Dotter_2016} . We used MESA code to calculate $ t_{ms} $ (see table \ref{tab:stars_params}). The results of total number of WIMPs are presented in figure \ref{fig:total_wimps}. It is concluded that

massive stars collect less total amount of WIMPs when they leave main-sequence. It is related to the fact that massive stars evolve faster and consequently they have lower $ t_{ms} $ in comparison with low-mass stars (see table \ref{tab:stars_params}). As a result DM affects low-mass stars structure more efficiently than high-mass stars. These results are in agreement with the results that are obtained by references \cite{Lopes2011, Zentner_2011, Casanellas_2009, Scott_2008} .

\begin{table*}[b]
	\caption{\label{tab:stars_params}
		Stellar parameters that are obtained using MESA code for stars with different stellar masses. $ t_{ms} $ is the main-sequence lifetime of each star. We followed the definitions in paper \cite{Dotter_2016} to calculate main-sequence lifetimes. Radius, central temperature and central density of each star is reported when stars are in the Intermediate Age Main Sequence (IAMS) phase (see paper  \cite{Dotter_2016} for extra explanation about the IAMS).
	}
	
	\begin{tabular}{m{5em}m{5em}m{5em}m{10em}m{8em}}
		\centering Mass of star $ \left(  M_{\odot}  \right) $ & \centering $ t_{ms} \newline (Giga Year) $ & \centering Radius  $  \left( R_{\odot} \right) $ & \centering Central Teprature  (Milion Kelvin) & central  density  $ \left( g \: cm^{-3} \right)$ \\
		\hline
		0.5 & 130.112 & 0.5042 & 9.922 & 157 \\
		1 & 11.438 & 1.0162 & 15.633 & 148 \\
		1.5 & 2.372 & 1.8889 & 21.038 & 114 \\
		2 & 1.019 & 2.3067 & 22.600 & 72 \\
		2.5 & 0.546 & 2.8596 & 24.406 & 55 \\
		3 & 0.334 & 3.0627 & 25.328 & 41 \\
	\end{tabular}
	
	\footnotetext[1]{When stars are in Intermediate Age Main-Sequence (IAMS) phase}.
\end{table*}

\subsection{\label{sec:sec2-3}WIMPs Distribution Inside Stars}

\begin{figure*}
	\includegraphics[width=\textwidth]{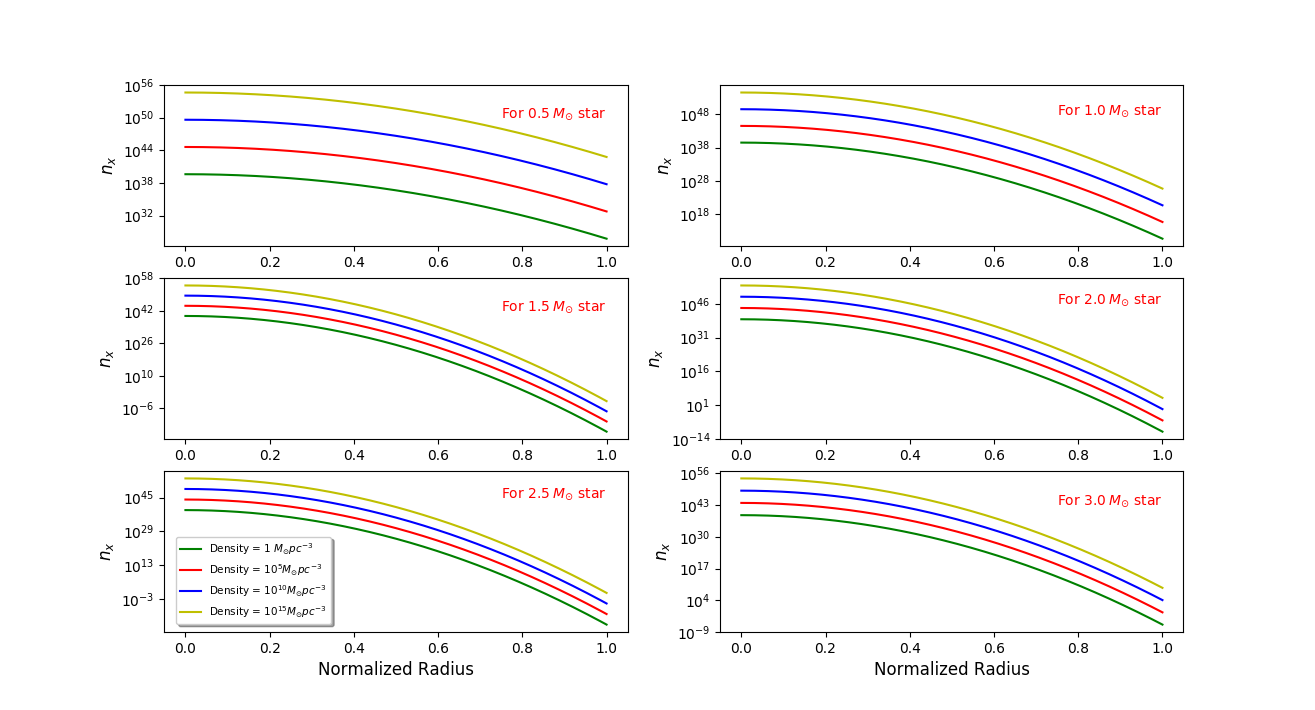}
	\caption{\label{fig:wimp_distru} (color online). Distribution of WIMPs inside stars with different stellar masses. Different colors represent different DM densities. For $ 0.5 M_{\odot} $ and $ 1 M_{\odot} $  stars, WIMPs distribution extends beyond the radius of the star while for others, WIMPs distributions are almost completely inside the stars. In all stars, WIMPs accumulate in central regions of the stars and reduce toward the outer region exponentially. So, as a result, during the evolution of all stars, WIMPs affect the physics of inner regions more efficiently than outer regions.}
\end{figure*}

The distribution of WIMPs inside stars is considered Boltzmannian \cite{Dearborn_1990} :

\begin{eqnarray} \label{eq:WIMPs_Distru}
n_{x}\left( r \right) = N_{x} \pi ^{-3/2} r_{x}^{-3} e^{ -r^{2}/ r_{x}^{2}  }
\end{eqnarray}

where $ N_{x} $ is the total number of WIMPs inside the star, $ T_{c} $ is the central temperature of the star and $ r_{x}^{2} = (3/2\pi) T_{c} \left(  m_{x} G_{N} \rho_{c}  \right) ^{-1} $ in which $ G_{N} $ is the Newtonian constant of gravitation and $ \rho_{c} $ is the central density of star. We used MESA code to obtain $ T_{c} $ and $ \rho_{c} $ at the Intermediate Age Main-Sequence (IAMS) phase \cite{Dotter_2016} . The results are presented in table \ref{tab:stars_params}. Using equation \ref{eq:WIMPs_Distru}, WIMPs distribution inside stars with different WIMP masses and different DM density environments are presented in figure \ref{fig:wimp_distru}. In figure \ref{fig:wimp_distru}, horizontal axes are normalized to the stars radius. We again used MESA code to obtain radii of stars with different stellar masses (see table \ref{tab:stars_params}).

Paying attention to figure \ref{fig:wimp_distru}, it is noticeable that, most of the WIMPs that are captured by the stars are accumulated near the central regions i.e. WIMPs affect the structure of the central regions more efficiently than outer regions. It is also seen from Figure \ref{fig:wimp_distru} that in some stars (for instance stars with masses $ 0.5 $ and $ 1 M_{\odot} $) WIMPs distribution extends beyond the stars radius, while in others (for instance $ 1.5 M_{\odot} $ and $ 2.0 M_{\odot} $ stars) WIMPs are distributed almost completely inside stars. This behavior is not a monotonic function of just mass of the stars, but it is a complex function of $ N_{x} $, $ \rho_{c} $ and $ T_{c} $ (according to equation \ref{eq:WIMPs_Distru}). This treatment is explaining why WIMPs affect low-mass stars structures and evolutionary tracks unlike the high-mass ones.

\subsection{\label{sec:sec2-4}Energy Produced Through WIMPs Annihilation}

By passing the time, the total number and local density of WIMPs inside stars will increase. Eventually, at the time $ t_{eq} $ stars reach to the equilibrium state, that is, the number of WIMPs that are capturing become equal to the number of WIMPs that are annihilating inside the stars. Multiplying produced energy through WIMP annihilation $ \left( 2m_{x} \: c^{2} \right) $ by capture rate, results the total luminosity that is produced through WIMPs annihilation. $ t_{eq} $ for main-sequence stars is \cite{Salati_1989} :

\begin{eqnarray} \label{eq:t_eq}
t_{equilibrium} \simeq \frac{109}{\rho_{6}^{1/2}} \left( \frac{\sigma_{\odot}}{\sigma_{s}} \right)^{1/2} \left( \frac{10^{-26}cm^{3}sec^{-1}}{\left\langle \sigma_{ann}v \right\rangle} \right) ^{1/2}\nonumber\\
\times \left( \frac{M}{M_{\odot}} \right)^{0.6} \left( \frac{m_{p}}{m_{x}} \right)^{1/4} \left( yr \right) \:\:\:\:\:\:\:\:\:\:\:\:\:\:\:\:\:\:\:\:\:\:\:\:
\end{eqnarray}

Luminosity that is produced through WIMPs annihilation after time $ t_{eq} $ will be \cite{Salati_1989}:

\begin{eqnarray} \label{eq:L_x}
\frac{L_{x}}{L_{\odot}} = 4 \times 10 ^{-5} \: \left( \frac{\rho_{x}}{1 \: M_{\odot} \: pc^{-3}} \right) \left( \frac{M}{M_{\odot}} \right) \: \left( \frac{R}{R_{\odot}} \right) \nonumber\\ \:\:\:\:\:\:\:\:
\times \left[ 1 + 0.16 \: \left( \frac{R}{R_{\odot}}\right)  \left( \frac{M_{\odot}}{M}\right) \right] \nonumber\\ \:\:\:\:\:\:\:\:\:\:\:\:
\times min \left[ 1 , \frac{\sigma _{s}}{\sigma_{\odot}}  \frac{M}{M_{\odot}}  \left(  \frac{R_{\odot}}{R} \right) ^{2}  \right]
\end{eqnarray}

\begin{figure*}
	\includegraphics[width=\textwidth]{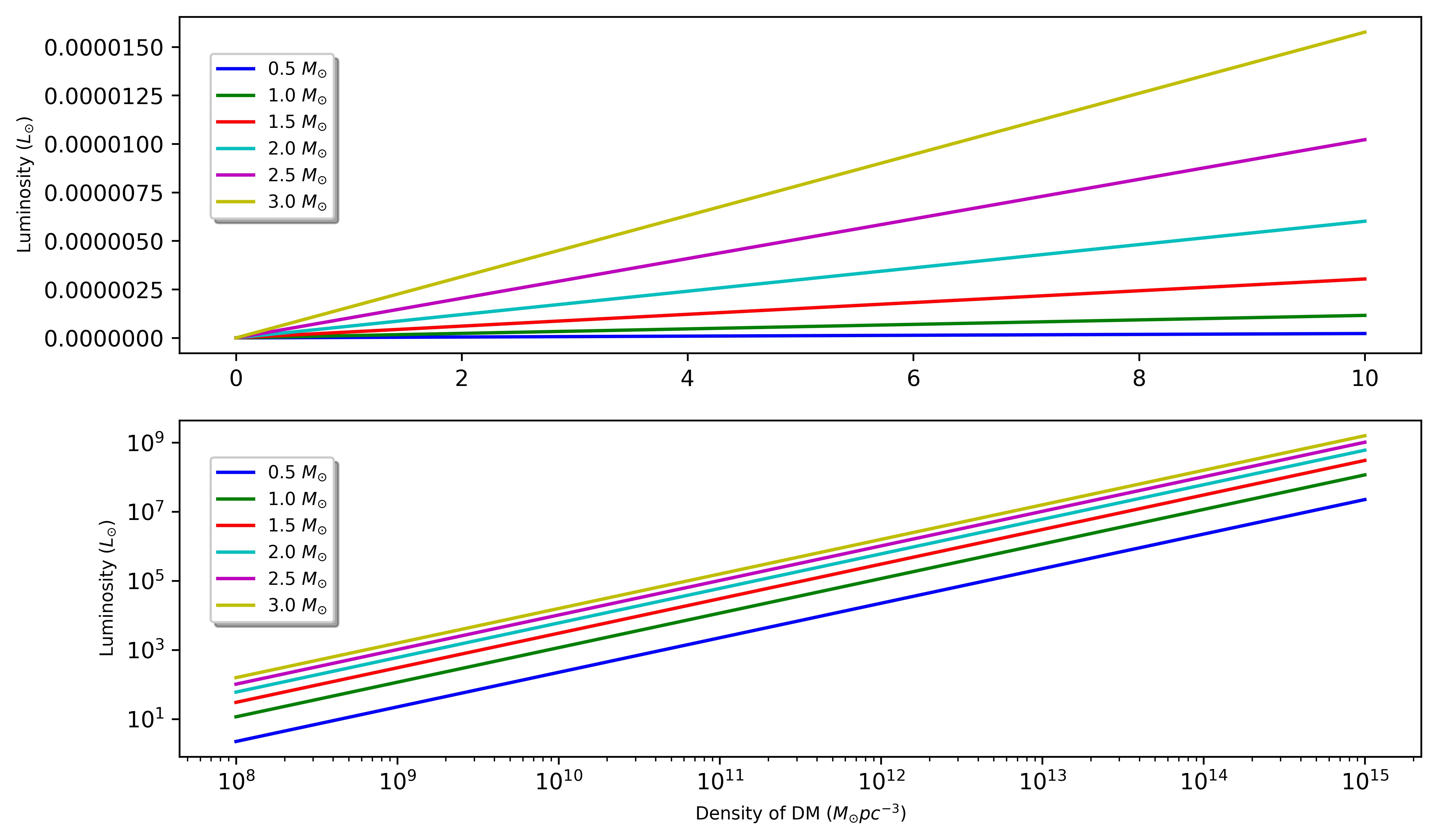}
	\caption{\label{fig:L_x} (color online). Luminosity produced through WIMPs annihilation inside stars as a function of DM density. Above part is for low DM density environments and its axises are linear. Below part is for high DM density environments and its axises are logarithmic. For a star at sun position in the Milky Way Galaxy (which DM density is about $\sim 8.16 \times 10 ^{-3} \: M_{\odot} / pc^{3}$ \cite{BERTONE_2005}) the portion of extra luminosity produced through WIMPs annihilation is about $ 10^{-8} L_{\odot}$. This amount is negligible in comparison with the amount of energy that is produced through baryonic matter energy production cycles. Additionally, near the Galactic MBH, where the density of DM is about $10^{15} \: M_{\odot} \: pc^{-3}$ , stars can produce $ \sim 10^{8} L_{\odot} $ energy through WIMPs annihilation.} 
\end{figure*}

By using equation \ref{eq:L_x} it is possible to calculate the produced energy through WIMPs annihilation in units of the Sun luminosity $ L_{\odot} $. The results are presented in figure \ref{fig:L_x} for different stellar masses. To illustrate detail for low DM density environments, we divided figure \ref{fig:L_x} to two parts. The above part is for low DM density environments (for instance for stars near the sun in the Milky Way Galaxy) which its horizontal axis is in linear scale. The below part devoted to high DM density environments (for instance for stars near the Galactic MBH) with horizontal axis in logarithmic scale.

It is found from Figure \ref{fig:L_x} that the luminosity of the sun that is produced through WIMPs annihilation (i.e $ L_{x} $) is negligible compared to the total luminosity that is produced through baryonic matter energy production cycles (i.e $ L_{\odot} $)(DM density near the sun position is about $ \simeq 8.16 \times 10 ^{-3} \: M_{\odot} / pc^{3} $ \cite{BERTONE_2005}). If a star with the solar mass evolves near Galactic MBH, where density of DM is about $\sim 10 ^{15} M_{\odot} pc ^{-3}$, it can produce luminosity of the order $ \sim 10^{8} \: L_{\odot} $ through WIMPs annihilation. It is clear that this new source of energy can not be ignored and must be considered in stellar evolutionary models.

CSEMs are not able to interpret the behaviour of the stars that are located near the Galactic MBH, then PYP arises. At this point, it is worth mentioning again the fact that, to solve the PYP one must consider all the unique and severe physical condition that exist near the Galactic MBH (as discussed in introduction section). In this work we just focused on DM effect and did not consider other special physical conditions for the reasons of the simplicity. 

\section{\label{sec:level4}Conclusion}

Effect of DM density on stars evolutions and structures is investigated. The results of numerical simulations show that DM does not affect low-mass stars and high mass stars in the same way. Although capture rate for high-mass stars is more than low-mass stars, but total number of WIMPs that are captured by low-mass stars is higher than that of the high-mass ones, when they leave main-sequence. The reason for this behavior is that low-mass stars spend more time on the main-sequence phase than high-mass ones. So, they have enough time to capture more WIMPs. We implemented new source of energy through WIMPs annihilation through  MESA stellar evolutionary code. It is found that this new source of energy causes stars to deviate from the standard evolutionary path on the H-R diagram i.e. the higher the DM density the higher the deviation.

It is showed that WIMPs distribution inside low-mass stars are different from high-mass ones. In fact, WIMPs distribution is not a monotonic function of just mass of the stars, but it is a complex function of parameters like $ N_{x} $, $ \rho_{c} $ and $ T_{c} $.

According to figure \ref{fig:L_x}, energy produced through WIMPs annihilation can be as large as $ \sim 10^{8} L_{\odot} $ for stars that are located near the Galactic MBH. To interpret the behavior of the strange stars that are evolving near the Galactic MBH one must consider this new source of energy while stars are evolving. By considering this new source of energy, stars follow different evolutionary pathes on the H-R diagram. When stars finish the source of energy that comes from baryonic matter, WIMPs annihilation play an important role in the total luminosity of the stars. At this phase, though the spectrum of the stars look like evolved and old stars (because of increasing the frequency of hevier elements), but their luminosities are like active and young stars. In other words, stars manifest signs of young and old stars simultaneously. As CSEMs do not consider WIMPs annihilation in their assumptions, the PYP arises in these models.

Our simulations show that in order to solve the PYP, one must consider all extreme and unique physical conditions that are present near the Galactic MBH. At present work, we considered only DM effect.

\section*{Acknowledgement}
Special thanks are due to Dr Amin Rezaei Akbarieh (From university of Tabriz, Iran) for useful and helpful discussions about DM related subjects, and Dr Mohammad Hosseinirad (from Ferdowsi University of Mashhad, Iran) for useful discussions about programming aspects of this research. Also, E. Hassani thanks Dr Anya Samadi (From university of Tabriz, Iran) and  Dr Ehsan Moravveji (from the Institute of Astronomy of KU Leuven, Belgium) for useful discussions during the research. The authors would like to thank Mesa-users team for satisfactory answers to research related questions. All figures were generated using Matplotlib, a 2D Graphics Environment library for python \cite{Hunter}.

\bibliographystyle{unsrt}
\bibliography{references}

\end{document}